\documentclass[preprint,preprintnumbers,amsmath,amssymb,floatfix,nofootinbib]{revtex4}
\usepackage{epsf, array, color}
\usepackage{graphicx}
\usepackage{subfigure}
\usepackage{bbold}

\newcommand{\bea}{\begin{eqnarray}}
\newcommand{\eea}{\end{eqnarray}}
\newcommand{\nn}{\nonumber \\}

\newcommand{\ep}{\epsilon}

\def\beq{\begin{equation}}
\def\eeq{\end{equation}}
\begin{document}

\title{Yukawa Corrections to Higgs Production in Top Partner Models}
\author{S.~Dawson$^{(1)}$}
\author{E.~Furlan$^{(1,2)}$}

\affiliation{
$^{(1)}$ Department of Physics, Brookhaven National Laboratory, Upton, New York, 11973
\\
$^{(2)}$ Fermi National Accelerator Laboratory, P.O. Box 500, Batavia, Illinois, 60510
\vspace*{.5in}}

\preprint{FERMILAB-PUB-13-476-T}
\date{\today}

\begin{abstract}
Higgs production from gluon fusion is sensitive to the properties of heavy colored fermions and 
to the Yukawa couplings, $\frac{Y_F M_F}{v}$,
of these particles to the Higgs boson.  We compute the two--loop, 
${\cal O}\left(\left(\frac{Y_F M_F}{v}\right)^3\right)$ 
contributions of new high mass fermions to 
Higgs production.  In the Standard Model, these contributions are part of the well-known 
electroweak corrections and are negligible. However, in models
with TeV scale fermions, such as top partner or composite models,  Yukawa corrections are 
enhanced by effects of
${\cal O}\left(\left(\frac{Y_F M_F}{v}\right)^3\right)$ 
and are potentially significant due to the large mass of the new quarks. 
We examine the size of these top partner 
Yukawa corrections to Higgs production for parameter choices which are allowed
by precision electroweak constraints. 
\end{abstract}

\maketitle

\section{Introduction}
The discovery of a 126~GeV Higgs boson leads to the question of whether this particle is the single 
scalar field predicted by the
Standard Model, or whether it is the remnant of some more complicated theory.  
Composite models~\cite{Agashe:2004rs,Contino:2006qr,Espinosa:2010vn}  
and models where the Higgs is a pseudo-Goldstone boson of a
broken symmetry such as Little Higgs models~\cite{ArkaniHamed:2002qy,Low:2002ws,Hubisz:2004ft,Chen:2003fm}
typically contain new heavy fermions which
are not present in the Standard Model.  These fermions can mix with the observed quarks and contribute
 to 
Higgs production and decay.   The properties of new charged $-{1\over 3}$
fermions which can mix with the Standard model $b$ quark are greatly restricted by measurements
of $Z\rightarrow b {\overline b}$ decays~\cite{Bamert:1996px,Freitas:2012sy} and so we will concentrate on
fermionic top partners which can mix with the Standard Model top quark.
 Heavy charged ${2\over 3}$ fermions have been searched for at the LHC, 
and depending on their decay modes,  are restricted to be heavier than $600-700$~GeV~\cite{cmstop}. 
The properties of these potential new heavy fermions are also strongly
constrained both by precision electroweak 
measurements~\cite{Picek:2008dd,Dawson:2012mk,Aguilar-Saavedra:2013qpa,Fajfer:2013wca,Grojean:2013qca,Csaki:2003si,Berger:2012ec,Anastasiou:2009rv,
Gori:2013mia}, 
and by the requirement that the Higgs production rate, $gg\rightarrow H$, be close
to the measured value~\cite{Dawson:2012di,Bonne:2012im,Azatov:2012bz,Espinosa:2012ir,Kearney:2012zi,Carmi:2012in}.  Precision measurements
of the Higgs production and decay rates offer a window into this possible new high scale physics,
and in this paper we focus on quantifying the predictions of top partner models.  

New heavy fermions can contribute to both $gg\rightarrow H$ and 
$ H\rightarrow\gamma\gamma$ and, because of the large top partner mass, $M_F$,
the  Higgs--fermion Yukawa couplings, $\frac{Y_F M_F}{v}$, may generate large 
${\cal O}\left(\left(\frac{Y_F M_F}{v}\right)^3\right)$ contributions at two loops.  
We compute the effects of the two--loop Yukawa couplings of top partners to  Higgs production
from gluon fusion using the low energy theorems valid in the 
$M_H \ll 2 M_F$ limit~\cite{Kniehl:1995tn,Low:2009di}. 
These corrections are part of the complete two--loop 
electroweak corrections to Higgs production from gluon fusion.  For the Standard Model, the Yukawa corrections
have been known for some time~\cite{Djouadi:1994ge,Djouadi:1997rj}
along with the complete two--loop electroweak
corrections~\cite{Aglietti:2004nj,Degrassi:2004mx,Actis:2008ts,Anastasiou:2008tj}. The full electroweak corrections are also known for 
a degenerate $4^{th}$ generation of heavy fermions which does not mix with the Standard
Model fermions~\cite{Passarino:2011kv}. The physical top mass, $m_t=173$~GeV, is not large enough for the Yukawa 
corrections to be the dominant contribution to the electroweak corrections, 
but for large top masses (say $m_t\sim 700$~GeV), the Yukawa corrections 
would become the most significant contribution to the two--loop
electroweak effect. Therefore, in
top partner models where the fermion mass is at the TeV scale, these 
${\cal O}\left(\left(\frac{Y_F M_F}{v}\right)^3\right)$ Yukawa corrections may be numerically significant.   

Technical details of our calculation  are contained in Section~\ref{sec:calc}.  We begin with a review 
of the low energy theorem as applied to the two--loop Yukawa corrections to Higgs production 
and include a discussion of renormalization and our technique
for expanding two--loop integrals. We demonstrate the validity of our techniques 
by reproducing the Standard Model (SM)  result for the Yukawa corrections to $gg\rightarrow H$ in Section \ref{sec:results}. 
Our new results are in Section \ref{sec_basics}, where we consider the class of models 
that contains a  top partner which is an $SU(2)_L$ singlet that mixes 
with the Standard Model top quark.
In Section \ref{sec:phen}, we discuss the relevance of our results for Higgs precision measurements
 and the search for 
new physics effects through the measurement of Higgs properties.

\section{Calculation techniques}
\label{sec:calc}
\label{sec:LET}
We are interested in the two--loop ${\cal O}\left(\left(\frac{Y_F M_F}{v}\right)^3\right)$ 
contributions to the gluon fusion production of a Higgs boson,
where $F$ is a heavy quark ($M_F \gg M_H/2$ ) coupling to the Higgs boson.
Since direct searches for top partners require $M_F > 700$~GeV~\cite{cmstop}, these contributions can potentially
give effects enhanced by powers of $M_F$.
The interactions of the heavy quarks with the Higgs boson are parametrized as,
\beq
	-{\cal L}_Y = \sum_F M_F^0 \left( 1 + Y_F^0 \frac{H^0}{v^0} \right) \overline{F}^0 F^0 \, ,
\label{eq:genHFF}
\eeq
where the superscript $0$ denotes the unrenormalized values of the parameters.
In the Standard Model, $Y_t = 1$. 

\subsection{Low Energy Theorem}
We use the low energy theorem to compute the leading contribution in ${M_H^2\over M_F^2}$ to
the $gg\rightarrow H$ process.
For a soft Higgs boson, $p_H \to 0$, the amplitude ${\cal A}_{gg \to H}$ is related 
to the gluon vacuum polarization amplitude, 
${\cal A}_{gg}=-i \Pi_{\mu \nu}^{AB} $~\cite{Ellis:1975ap, Shifman:1979eb, Spira:1995rr, Kniehl:1995tn},
\beq
\label{eq:ggHLET}
	\lim_{p_H \to 0} {\cal A}_{gg \to H} = 
	\frac{1}{v^0} \sum_F Y_F^0 M_F^0 \frac{i \partial}{\partial M_F^0} {\cal A}_{gg} \;.
\eeq
This is equivalent to inserting an additional heavy-quark propagator with the emission 
of a zero-momentum Higgs boson. (We have assumed that we are working with
the  quark mass eigenstates). The differentiation is performed on the bare masses 
coming from propagators, while mass-dependent couplings are to be treated as 
constants. The renormalization is performed after taking the derivatives.  It is 
straightforward to extend this approach to loop corrections to Higgs 
production~\cite{Spira:1995rr,Chetyrkin:1997iv,Chetyrkin:1997un,Harlander:2002wh,Anastasiou:2002yz,Dawson:1998py,Dawson:1990zj}.

At one loop, the gluon polarization tensor $\Pi^{\mu \nu}_{AB}(p^2)$ is 
\bea
	\Pi^{\mu\nu~1L}_{AB}(p^2)
	&=&
	{\alpha_s^0\over \pi} 
	\delta_{AB} \left(g^{\mu\nu} p^2-p^{\mu} p^{\nu}\right)
	[N] \sum_F \left\{
		(M_F^0)^{-2 \ep} 
		\biggl[{1\over 6\epsilon} + {\cal O} \biggl(\frac{p^2}{(M_F^0)^2}, \ep \biggr) \biggr] \right\} \, .
%		-I_3\biggl({p^2\over m^2}\biggr)\biggr]\, .
%	\nonumber \\
	\;
\end{eqnarray}
%where, $I_1(a)\equiv \int_0^1 dx x (1-x)\ln(1-ax(1-x))$
The amplitude for $gg\rightarrow H$ from Eq.~\ref{eq:ggHLET} is,
\begin{eqnarray}
\label{eq:AggH1L}
	{\cal A}_{gg \to H}^{1L}&=&
	- {\alpha_s^0\over 3 \pi v^0}  
	\delta_{AB} \left(g^{\mu\nu} p^2-p^{\mu} p^{\nu}\right) [N]
	\sum_F Y_F^0 (M_F^0)^{-2 \ep}
	\nonumber \\
	&\equiv & \Sigma_F {\cal A}^{1L,F}_{gg\rightarrow H} \, ,
\eea
where,
\beq
\label{eq:normNQ}
	[N] = \Gamma(1+\ep) (4 \pi \mu^2)^\ep
	\xrightarrow{\ep \to 0} 1 \;.
\eeq
and the sum is over all heavy fermions. 

\subsection{Techniques for two--loop integrals}
We are interested in the two--loop contributions to the gluon two--point function which are 
enhanced by powers of 
the Yukawa couplings and so we neglect the ${\cal{O}}(g^2)$ contributions from $W$ and $Z$ exchange. 
In Landau gauge, the Goldstone bosons are massless and couple with Yukawa strength to the massive
fermions, and so are included in the calculation. The ${\cal O}\left(\left(\frac{Y_F M_F}{v}\right)^3\right)$ 
contributions form a gauge invariant
subset of the complete two--loop electroweak corrections.  

Each of the diagrams has a contribution
of the form (for external momentum $p$), 
\begin{equation}
\Pi_i^{\mu\nu}(p^2) =a_i g^{\mu\nu}+b_i p^\mu p^\nu\, .
\end{equation}
Gauge invariance requires that $\sum_i b_i=0$.
The coefficients are found by taking contractions 
with $g^{\mu\nu}$ and $p^\mu p^\nu$:
\begin{eqnarray}
a_i&=& \frac{1}{d-1} \Pi_i^{\mu\nu}(p^2)  \left( g_{\mu\nu} - \frac{p_\mu p_\nu}{p^2} \right)  \; ,\\
b_i&=& -\frac{1}{p^2 (d-1)} \Pi_i^{\mu\nu}(p^2) \left( g_{\mu\nu} - d\frac{p_\mu p_\nu}{p^2} \right)  \; ,
\label{eq:projdef}
\end{eqnarray}
with $d=4-2\epsilon$.

The strategy is to expand the loop integrals in powers of external momentum over the heavy 
mass scale in the loop, $M_F$. The numerators of the integrals have the form,
\begin{equation}
(k_1\cdot  p)^j  (k_2\cdot p)^m  \times ({\hbox{powers~of~}}k_1^2,k_2^2, k_1\cdot k_2)\; ,
\end{equation}
where $k_1, k_2$ are the loop momenta. These integrals 
can be symmetrized using the techniques in the appendix of Ref.~\cite{Dawson:1993qf} .  

In the limit where all the fermions in the loop are much heavier than the external mass scale 
(which will generically be of
${\cal O}(p^2 \sim M_H^2)$), we can calculate the two--loop 
integrals by expanding in powers of ${p^2\over M_F^2}$ and retaining the leading term. 
Due to the small--momentum expansion, the integrals that we need to compute are all 
two--loop vacuum bubbles. If the Higgs and Goldstone boson interactions do not mix quarks with the 
same quantum number, as in the Standard Model and its four--generation extension, the 
vacuum bubbles only depend on one heavy mass scale, $M_F$. Their general form is 
\begin{equation}
	B \left(M_F,M_F,0;n_1,n_2,1\right) =\int {d^d k_1\over (2\pi)^d}
	\int {d^d k_2\over (2\pi)^d} 
	{1\over (k_1^2-M_F^2)^{n_1} (k_2^2-M_F^2)^{n_2} (k_1+k_2)^2}\, .
\end{equation}
Explicit expressions for these integrals are given in 
Refs.~\cite{Hoogeveen:1985tf,Dawson:1993qf,Dawson:1992cy,Davydychev:1995mq}.  Alternatively, one 
can use integration by part identities to reduce these integrals to the master 
integral~\cite{Hoogeveen:1985tf,Dawson:1993qf,Dawson:1992cy,Davydychev:1995mq}
\bea
	B\left(M_F,M_F,0;1,1,1\right) &=& - \frac{M_F^{2-4\ep}}{(4 \pi)^4} [N]^2
		\left({1\over \ep^2}+{3\over \ep} + 7\right) \; .
\eea
We obtain these relations with the program AIR~\cite{Anastasiou:2004vj}. 

If more quarks with the same quantum numbers are present, and the Higgs and Goldstone 
boson interactions mix them, we also have  the ``off--diagonal" contribution where both the heavy 
quarks plus the boson run in the loops. We need the additional 
two--loop,  two--masses scalar master integral, 
\beq
	B \left(M_F,M_{F'},0;1,1,1\right) = \int {d^d k_1 \over (2 \pi)^d} {d^d k_2 \over (2 \pi)^d} 
	{1 \over k_1^2-M_F^2} {1 \over k_2^2-M_{F'}^2} {1 \over (k_1+k_2)^2} \;,
\eeq
where $M_F, M_{F'}$ are the masses of the two heavy quarks. 
In the literature the integral with three massive lines is known~\cite{Davydychev:1995mq}, 
\bea
\label{eq:MI_davyd}
	B \left(M_F,M_{F'},m;1,1,1\right) &=& \int {d^d k_1 \over (2 \pi)^d} {d^d k_2 \over (2 \pi)^d} 
	{1 \over k_1^2-M_F^2} {1 \over k_2^2-M_{F'}^2} {1 \over (k_1+k_2)^2-m^2} 
\nn &=&	
	 {1 \over 2} {m^{2-4\ep} \over (4\pi)^4} 
	{[N]^2 \over (1-\ep) (1-2\ep) } 
	\left\{ 
		- {1 \over \ep^2}  (1+x+y) + {2 \over \ep}  (x \ln x + y \ln y)
	\right. 	\nn && 
	\left.
		- x \ln^2 x - y \ln^2 y + (1-x-y) \ln x \ln y
		- \lambda^2(x, y) \Phi^{(1)}(x,y) \right\} %+ {\cal O}(\ep)  
		\; . \nn
\eea
The functions $\lambda^2(x, y)$ and $\Phi^{(1)}(x,y)$ are 
\bea
	\lambda(x,y) &=& \sqrt{(1-x-y)^2 - 4 x y} \; , \nn
	\Phi^{(1)}(x,y) &=& {1 \over 2 \lambda} \left\{ 
		4\, \mbox{Li}_{2}(1-z_1) + 4\, \mbox{Li}_{2}(1-z_2) + 4\, \mbox{Li}_2(1-z_3)
		\right. \nn 
		&& \left.
         + \ln^2 z_1  + \ln^2 z_2 + \ln^2 z_3
         + 2 \ln x \ln z_1 + 2 \ln y \ln z_2 \right\} \; ,
\eea
with 
\beq
	z_1 = {(\lambda + x - y - 1)^2 \over 4y} \, , \quad
	z_2 = {(\lambda + y - 1 - x)^2 \over 4x} \, , \quad
	z_3 = {(\lambda + 1 - x - y)^2 \over 4xy} \;,
\eeq
and $x=M_F^2/m^2, y=M_{F'}^2/m^2$. 
Using this result, we compute the two--loop gluon self energy retaining the dependence 
on all three masses. We then take the limit $m \to 0$.

The virtual two--loop results for $g\rightarrow g$ depend on the specific model and will be given later. 
The two--loop contributions to $gg\rightarrow H$ from the heavy fermion loops are then found by
applying the low energy theorem of Eq.~\ref{eq:ggHLET}.

\subsection{Renormalization} 
Renormalization  of the $gg\rightarrow H$ amplitude requires the quark mass and wave function 
counterterms, the Higgs wave function counterterm,  and  the $F{\overline F} g$ and  
$F {\overline F}H$  vertex counterterms. 
The quark wave function renormalization, $Z_{2,F}$,  cancels against other counterterms 
and we do not need to compute it explicitly.
We briefly review the renormalization of the quark mass.  
We start from the bare Lagrangian, 
\beq
	{\cal L} =% -\frac{1}{4} G^{0,a}_{\mu \nu} G^{0, \mu \nu}_a +
	{\overline F}^0(i\partial \!\!\!/ -M_F^0)F^0 
	-g_s^0  {\overline F}^0  \gamma^\mu t^a G^{0, a}_\mu F^0 \;.
\eeq
The superscript $``0"$ denotes bare fields and couplings which are related to the renormalized 
ones by the renormalization constants, 
\beq
	\begin{array}{rclcrclcl}
		G^{0,a}_\mu&=&\sqrt{Z_3} G^a_\mu &\; , 
		&\quad
		F^0 &=& \sqrt{Z_{2,F}}F &=& \left(1+{\delta Z_{2,F}\over 2}\right) F \;, \\
		g_s^0 &=& \frac{Z_1}{Z_{2,F} \sqrt{Z_3}} g_s &\;,
		&\quad
		M_F^0 &=& Z_M M_F &=& \left(1 + {\delta M_F \over M_F} \right)  M_F \;.
\end{array}
\label{eq:def_counterterms}
\eeq
With these conventions the $F{\overline F} g$ vertex is renormalized by $Z_1$ and due to  
the Ward identities, $Z_1 = Z_{2,F}$. The quark propagator counterterm is, 
\beq
	\delta_F^{ct} = i \left[ (p \!\!\! /-M_F) \delta Z_{2,F} - \delta M_F \right] \;.
\eeq
We require the renormalized quark propagator $-i \Sigma(M_F, p\!\!\! /)$ (including the counterterms) 
to be canonically normalized and to have a pole at the renormalized mass, 
\beq 
	\Sigma(M_F, p\!\!\! / = M_F) = 0 \; , \qquad
	\Sigma'(M_F, p\!\!\! /) \! \!\mid_{p\!\!\! / = M_F} = 0 \; ,
\eeq
yielding, at one loop,
\beq
	\delta M_F = - \Sigma_{1L}(M_F, p\!\!\! /=M_F) \; , \qquad
	\delta Z_{2,F} = \Sigma_{1L}'(M_F, p\!\!\! /) \!\! \mid_{p\!\!\! / = M_F} \, ,
\eeq
where the sum of all the one--loop one--particle irreducible (1PI) insertions into the quark propagator 
is denoted as $-i \Sigma _{1L}(M_F, p\!\!\! /)$.

We now turn to the $F{\overline F} H$  vertex counterterm. The interaction of a fermion $F$ 
with the Higgs boson is 
\begin{equation}
\label{eq:mass_lag_for_renorm_2}
	{\cal L}_Y = 
%-{1\over 4} F_{\mu\nu}^{0,A}F^{0\mu\nu,A}
%+ {\overline F}^0(i\partial -g_s^0T^A G^{0 \mu}\gamma_\mu -m_F^0)F^0
	- \frac{H^0}{v^0} Y_F M^0_F {\overline F}^0 F^0 
	= - {g^0\over 2 M_W^0} H^0  Y_F M^0_F {\overline F}^0 F^0  \; .
\end{equation}
The $Y_F$ coupling and $g$ receive no ${\cal{O}}(Y_F^2)$ renormalization to the order in which 
we are working~\cite{Djouadi:1997rj,Kniehl:1994ju}. 
We introduce the renormalization constants, 
\beq
	\begin{array}{rclcrcl}
		H^0&=&\sqrt{Z_H}H=\left(1+{\delta Z_H\over 2}\right)H &\; , \quad
		(M_W^0)^2 &=& M_W^2\left(1+{\delta M_W^2\over M_W^2}\right) &\;.
	\end{array}
%	F^0=&\sqrt{Z_2^F}F= \left(1+{\delta Z_2\over 2}\right)\nonumber \\
%	G^{0,A}&=& \sqrt{Z_3}G^A\nonumber \\
%	H^0&=&\sqrt{Z_H}H=\left(1+{\delta Z_H\over 2}\right)H \;.
%\end{eqnarray}
 \eeq
%There is no contribution of order ${\cal O} \left(M_F^2/M_W^2\right)$ from the renormalization of $g$. 
In terms of the renormalized quantities, Eq.~\ref{eq:mass_lag_for_renorm_2} becomes  
\bea
	{\cal L} &=& 
	- {g\over 2 M_W} H  Y_F M_F {\overline F} F \left(1 +  {\delta M_F \over M_F} + \delta Z_{2,F} 
		+ \delta_3 \right) \; ,
\label{eq:renlag}
\eea
with 
\beq
	\delta_3 = \left( {\delta Z_H\over 2} - \frac{1}{2}{\delta M_W^2\over M_W^2}\right) \; .
\label{d3def}
\eeq

The Higgs wave function renormalization is computed from the sum of all 
1PI insertions into the Higgs propagator, 
$ - i \Pi_H(p^2)$,
%. Summing over all their contributions, the full Higgs propagator 
%in the bare theory reads
%\bea
%	\frac{i}{p^2-\left( m_H^0 \right)^2 - \Pi_0(p^2)} 
%	&=& \frac{1}{1-\Pi'(p^2)\vert_{p^2 = (m_H^0)^2}} 
%		\frac{i}{p^2-\left( m_H^0 \right)^2 - 
%				\frac{\Pi_0(p^2 = (m_H^0)^2)}{\Pi'(p^2)\vert_{p^2 = (m_H^0)^2} }
%				} \;.
%\eea
%In the second line we expanded around the Higgs mass. Therefore 
\beq
Z_H = \left[1-\Pi'_H(p^2)\vert_{p^2 = (M_H^0)^2}\right]^{-1} \;,
\eeq
which at one loop order yields 
\beq
	\delta Z_H = \Pi'_H(p^2)\vert_{p^2 = (M_H^0)^2} \;.
\eeq 
%Since we are adopting the $\overline{MS}$ scheme for the renormalization of the 
%electroweak sector, we can set the Higgs boson mass to zero in $\delta Z_H$. 
%The mass of the 
%heavy quarks in the loops already acts as an infrared (IR) regulator, ensuring that 
%all the poles are ultraviolet (UV). 
Similarly, the $W$ mass renormalization can be computed from the sum of all 
one-particle irreducible (1PI) insertions into the 
$W$ propagator.

We now combine these result to obtain the two--loop 
${\cal O}\left(\left(\frac{Y_F M_F}{v}\right)^3\right)$ 
counterterms for the $gg \to H$ 
amplitude in the limit $M_H <<2  M_F$. We have
\begin{itemize}
	\item from the quark mass counterterms on the internal legs, 
		\beq 
			{\cal A}_{M_F}^{2L, \,ct} = \sum_F \left[
				\left(i {\partial {\cal A}^{1L, F}_{gg \to H} \over \partial M_F} \right)
				\left( -i \delta M_F \right) \right] =  \sum_F \left[
				 \left( M_F {\partial {\cal A}^{1L, F}_{gg \to H}  \over \partial M_F} \right)
				 \left( {\delta M_F \over M_F} \right) \right] \; .
		\eeq
		The derivative of the one--loop amplitude with respect to the mass of the fermion 
		gives an extra fermion propagator, upon which we insert the mass counterterm. 
		We emphasize that the mass counterterm for the fermion $F$ needs to be inserted  
		only upon the one--loop amplitude containing that fermion. 
		As in the low energy theorem of Eq.~\ref{eq:ggHLET}, the derivatives only act on mass terms coming from 
		the internal propagators, and not on the masses from the Yukawa interactions. 
		We should notice that in the result, Eq.~ \ref{eq:AggH1L}, there is cancellation between 
		a mass from the Yukawa vertex, $Y_F{M_F^0 \over v^0}$, and a mass from the 
		propagator. With this in mind, 
		\beq
			M_F {\partial {\cal A}^{1L, F}_{gg \to H}  \over \partial M_F} 
			=
			-(2 \ep + 1 ) {\cal A}^{1L, F}_{gg \to H} \; ,
		\eeq
	\item from the quark wave function renormalization,
		\beq 
			{\cal A}_{Z_{2,F}}^{2L, \,ct} = 
				3 \sum_F \left[ \left(i {\cal A}^{1L,F}_{gg \to H} \right) \left( i \delta Z_{2,F} \right) \right] =
				- 3 \sum_F \left[\delta Z_{2,F} \, {\cal A}^{1L,F}_{gg \to H} \right] \; ,
		\eeq	
	\item from the $F{\overline F} g$ vertex counterterm, 
		\beq
			{\cal A}_{Z_1}^{2L, \,ct} = 
				2 \sum_F \left[ \delta Z_{2,F} \, {\cal A}^{1L,F}_{gg \to H} \right] \; ,
		\eeq
	\item from the $F{\overline{F}} H$  vertex of Eq.~\ref{eq:renlag}, 
		\beq
			{\cal A}_{Z_Y}^{2L, \,ct} = 
				\sum_F \left[
					{\cal A}^{1L,F}_{gg \to H} \left(  {\delta M_F \over M_F} + \delta Z_{2,F} + \delta_3 
					\right) \right] \; .
		\eeq
\end{itemize}
Combining these results we obtain 
	\bea
		{\cal A}^{2L, \,ct}_{gg \to H} &=& 
%			M_F {\partial {\cal A}^{1L}  \over \partial M_F} {\delta M_F \over M_F}
%			+ {\cal A}^{1L} \left(  {\delta M_F \over M_F} +\delta_3 \right) \; .
			\sum_F \left\{ \left[-(2 \ep + 1) {\delta M_F \over M_F}
			+\left(  {\delta M_F \over M_F} +\delta_3 \right) \right] {\cal A}^{1L,F}_{gg \to H} \right\}
		\label{eq:2L_renorm_interm}\\
		&=&
		 \sum_F \left[
		 	\left( -2 \ep  {\delta M_F \over M_F}  + \delta_3 \right) {\cal A}^{1L,F}_{gg \to H} 
		 \right]\;.
		\label{eq:2L_renorm}
	\eea
Since the one--loop result is finite, the counterterm receives a finite contribution from the 
pole of the quark mass renormalization and divergencies in the counterterm 
can only come from $\delta_3$. 

%The last two terms in Eq.~(\ref{eq:2L_renorm_interm}) arise from the 
%renormalization of the external Higgs and the input parameters, ${g\over M_W}$. The first  term 
%is necessary because
%the one-loop gluon vacuum polarization has been computed in the bare theory. 
%Alternatively~\cite{Adler:1976zt,Djouadi:1994ge,Kniehl:1995tn}, one can start from 
%the renormalized one loop self energy. In this case, the low energy theorem~(\ref{eq:ggHLET}) is 
%recast into the form 
%\bea
%	M_F^0 \frac{\partial}{\partial M_F^0} 
%	&=& {1 \over 1+ \gamma_M}  M_F \frac{\partial}{\partial M_F} \nn
%	&\simeq & (1- \gamma_M)  M_F \frac{\partial}{\partial M_F} \; ,
%\eea
%where 
%	\begin{equation}
%		\gamma_M = M_F {\partial \over \partial M_F} \log Z_M
%	\end{equation}
%is the mass anomalous dimension. 
%Substituting the definition of $Z_M$~(\ref{eq:def_counterterms}) and discarding  surface 
%integral terms, one indeed recovers the piece coming from the one-loop renormalization in 
%Eq.~(\ref{eq:2L_renorm}). We will use this as a cross check for the renormalization.  ***
%STILL NEED TO SHOW THIS AND DEMONSTRATE EXPLICITELY THAT WE GET EQ.\ref{eq:2L_renorm}***

\section{Results}
\label{sec:results}
\subsection{Standard Model}
As a check of our technique, we reproduce the well known ${\cal{O}}(\frac{m_t^3}{v^3})$ contributions 
to the gluon two--point function and to the $gg\rightarrow H$ 
amplitude in the limit $M_H\rightarrow 0$~\cite{Djouadi:1997rj,Djouadi:1994ge}.
We compute, for each diagram, the contractions with 
$g^{\mu\nu}$ 
and $p^\mu p^\nu$, which are shown in Table~\ref{tab:individual_tensor_contractions_SM}. 
The Standard Model with a massless $b$ quark  corresponds to $m_b = Y_b =0$,  $Y_t = 1$, 
and as a shorthand notation  we define $\frac{m_t}{v} \equiv y_t$.  A massless $b$ quark first
enters at two--loops.  The terms of ${\cal O}({p^4\over m_t^4})$ do not enter into our final results,
but are included as a check of our method and demonstration of gauge invariance.  

\begin{table}[]
\begin{center}
\begin{tabular*}{1\textwidth}{|m{0.25\textwidth}|m{0.365\textwidth}|m{0.355\textwidth}|}
\hline
	&$g_{\mu \nu}$ contraction& $p_{\mu} p_{\nu}/p^2$ contraction\\
\hline
\begin{center}\includegraphics[bb= 38 685 328 740, scale=0.39]{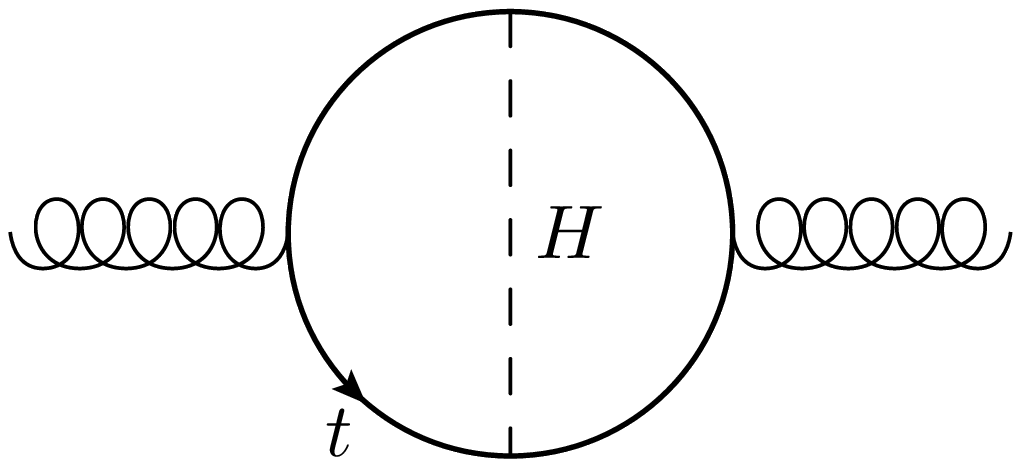} 
\end{center}
&
	\begin{tabular}{c}
		${\cal F}_t \left[1+\frac{3}{\ep}
		- \frac{p^2}{m_t^2} \left( \frac{157}{72} -\frac{1}{12 \ep} + \frac{1}{4 \ep^2}\right) \right.$
		 \\
		$	\left. - \frac{p^4}{m_t^4} \left( \frac{509}{1350}+\frac{1}{10 \ep} \right) \right]$
	\end{tabular}
&
	\begin{tabular}{c}$
		{\cal F}_t \left[ \frac{5}{8} +\frac{3}{4 \ep}
		+ \frac{p^2}{m_t^2} \left( \frac{1}{48} -\frac{1}{24 \ep} \right) \right.$ 
		\\
	$\left.	- \frac{p^4}{m_t^4} \frac{1}{240}\right]
$
	\end{tabular}
\\	\hline
\begin{center}\includegraphics[bb= 40 472 320 584, scale=0.39]{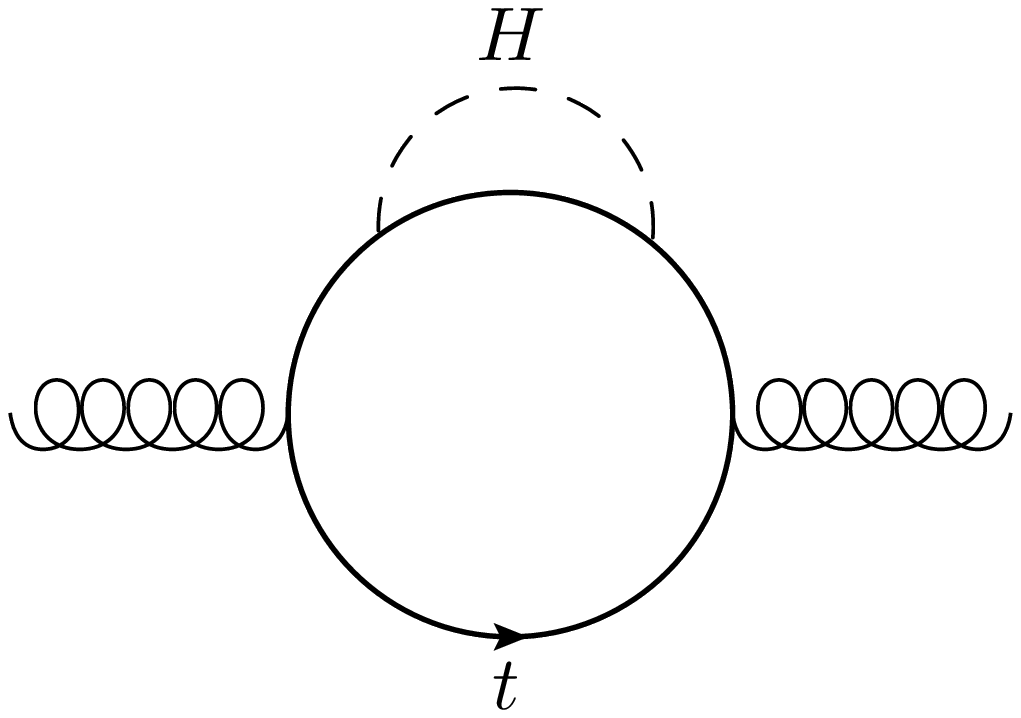} 
\end{center}
&
	\begin{tabular}{c}
		${\cal F}_t \left[ -\frac{1}{2} - \frac{3}{2 \ep} 
		+ \frac{p^2}{m_t^2} \!
		\left( \frac{161}{288} -\frac{29}{48 \ep} + \frac{1}{8 \ep^2}\right) \right.$
		 \\
		$	\left. - \frac{p^4}{m_t^4} \left( \frac{11}{75}+\frac{1}{10 \ep} \right) \right]$
	\end{tabular}
&
	\begin{tabular}{c}$
		{\cal F}_t \left[ -\frac{5}{16} -\frac{3}{8 \ep} 
		- \frac{p^2}{m_t^2} \left( \frac{1}{96} -\frac{1}{48 \ep} \right) \right.$ 
		\\
	$\left.	+ \frac{p^4}{m_t^4} \frac{1}{480}\right]
$
	\end{tabular}
\\ \hline
\begin{center}\includegraphics[bb= 470 685 760 735, scale=0.39]{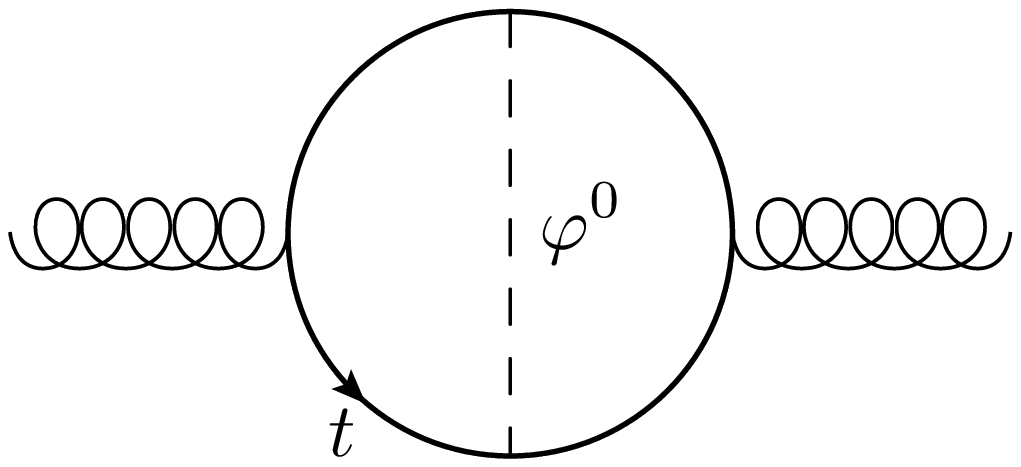} 
\end{center}
&	\begin{tabular}{c}
		${\cal F}_t \left[3+\frac{1}{\ep}
		+ \frac{p^2}{m_t^2} \left( \frac{5}{24} +\frac{1}{12 \ep} - \frac{1}{4 \ep^2}\right) \right.$
		 \\
		$	\left. + \frac{p^4}{m_t^4} \left( \frac{19}{180}-\frac{1}{10 \ep} \right) \right]$
	\end{tabular}
&
	\begin{tabular}{c}$
		{\cal F}_t \left[ \frac{7}{8} +\frac{1}{4 \ep}
		- \frac{p^2}{m_t^2} \left( \frac{5}{144} +\frac{1}{24 \ep} \right) \right.$ 
		\\
	$\left.	- \frac{p^4}{m_t^4} \frac{1}{144}\right]
$
	\end{tabular}
\\ \hline
\begin{center}\includegraphics[bb= 470 470 760 585, scale=0.39]{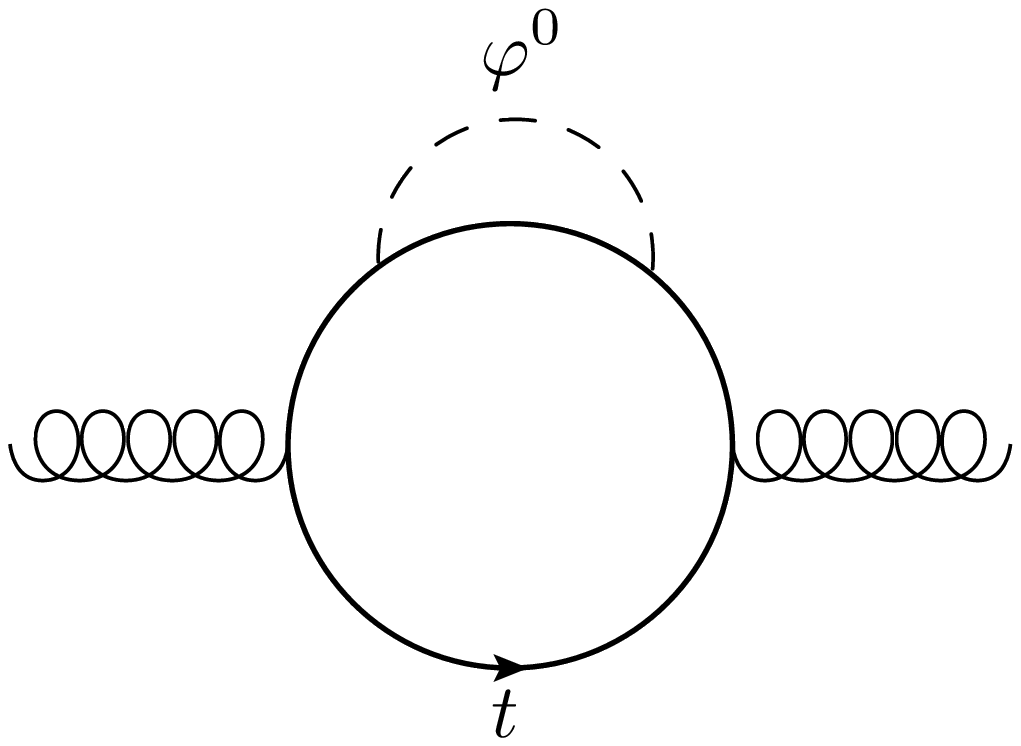} 
\end{center}
&	
\begin{tabular}{c}
		${\cal F}_t \left[ -\frac{3}{2} - \frac{1}{2 \ep} 
		+ \frac{p^2}{m_t^2} \! \left( -\frac{5}{96} \! +
			\! \frac{19}{48 \ep} \! + \! \frac{1}{8 \ep^2}\right) \right.$
		 \\
		$	\left. + \frac{p^4}{m_t^4} \left( \frac{1}{18}+\frac{1}{10 \ep} \right) \right]$
	\end{tabular}
&
	\begin{tabular}{c}$
		{\cal F}_t \left[ -\frac{7}{16} -\frac{1}{8 \ep} 
		+ \frac{p^2}{m_t^2} \left( \frac{5}{288} +\frac{1}{48 \ep} \right) \right.$ 
		\\
	$\left.	+ \frac{p^4}{m_t^4} \frac{1}{288}\right]
$
	\end{tabular}
\\ \hline
\begin{center}\includegraphics[bb= 45 290 320 335, scale=0.39]{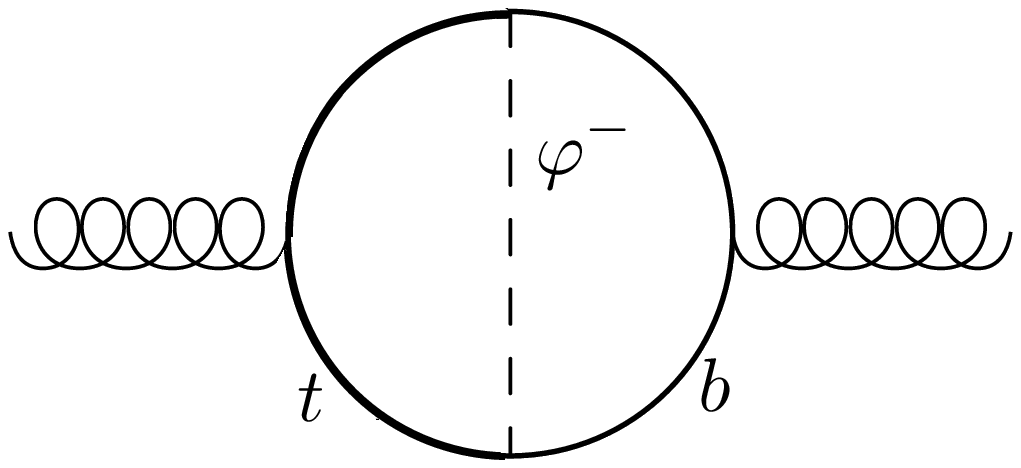} 
\end{center}
&
\begin{tabular}{c}
		${\cal F}_t \left[2+\frac{1}{\ep}
		- \frac{p^2}{m_t^2} \left( \frac{3}{16} -\frac{7}{24 \ep} \right) \right.$
		 \\
		$	\left. + \frac{p^4}{m_t^4} \left( \frac{167}{2160}+\frac{17}{360 \ep} \right) \right]$
	\end{tabular}
&
	\begin{tabular}{c}$
		{\cal F}_t \left[ \frac{5}{8} +\frac{1}{4 \ep}
		- \frac{p^2}{m_t^2} \left( \frac{11}{144} +\frac{1}{24 \ep} \right)  \right.$ \\
		$ \left. -  \frac{1}{72} \frac{p^4}{m_t^4}\right] $
	\end{tabular}
\\ \hline
\begin{center}\includegraphics[bb= 725 220 1010 330, scale=0.39]{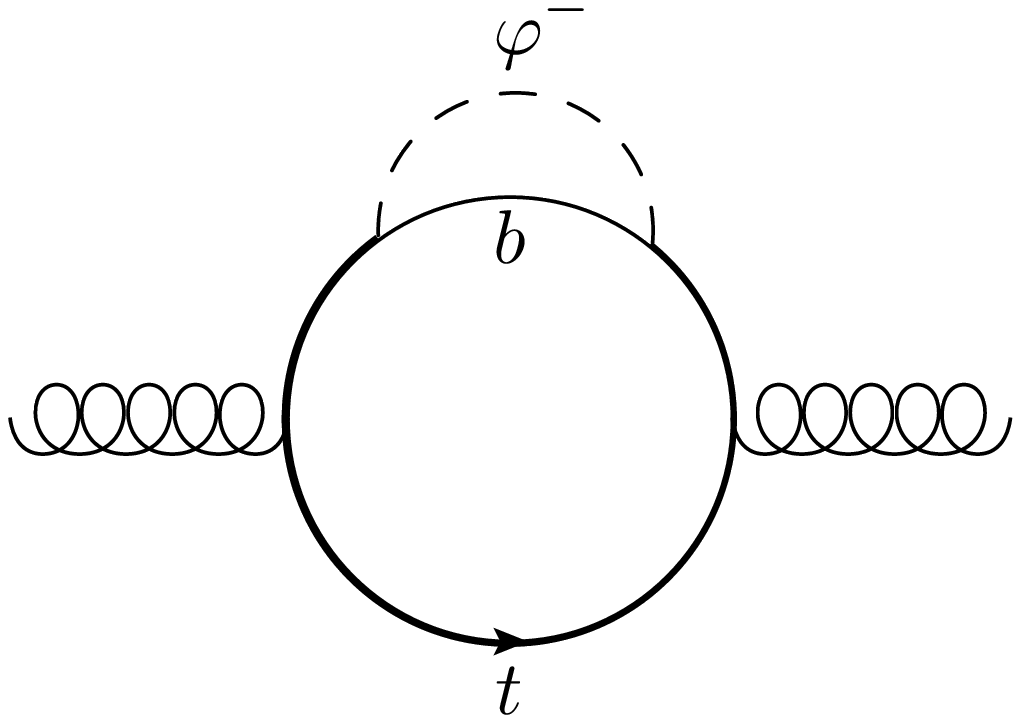} 
\end{center}
&	
\begin{tabular}{c}
		${\cal F}_t \left[ -1 - \frac{1}{2 \ep} 
		+  \frac{p^2}{m_t^2} \left( \frac{7 + 4 \pi^2}{96} -\frac{5}{48 \ep} \right. 
		\right.$
		 \\
		$	\left. \left. + \frac{1}{8 \ep^2}\right)  - \frac{7}{180}\frac{p^4}{m_t^4} \right] $
		 \\
	\end{tabular}
&
	\begin{tabular}{c}
		${\cal F}_t \left[ -\frac{5}{16} - \frac{1}{8 \ep} 
		+  \frac{p^2}{m_t^2} \left(\frac{11}{288} + \frac{1}{48 \ep}\right) \right.$\\
		$ \left. + \frac{1}{144}  \frac{p^4}{m_t^4} \right] $	
	\end{tabular}
\\ \hline
\begin{center}\includegraphics[bb= 380 220 670 335, scale=0.39]{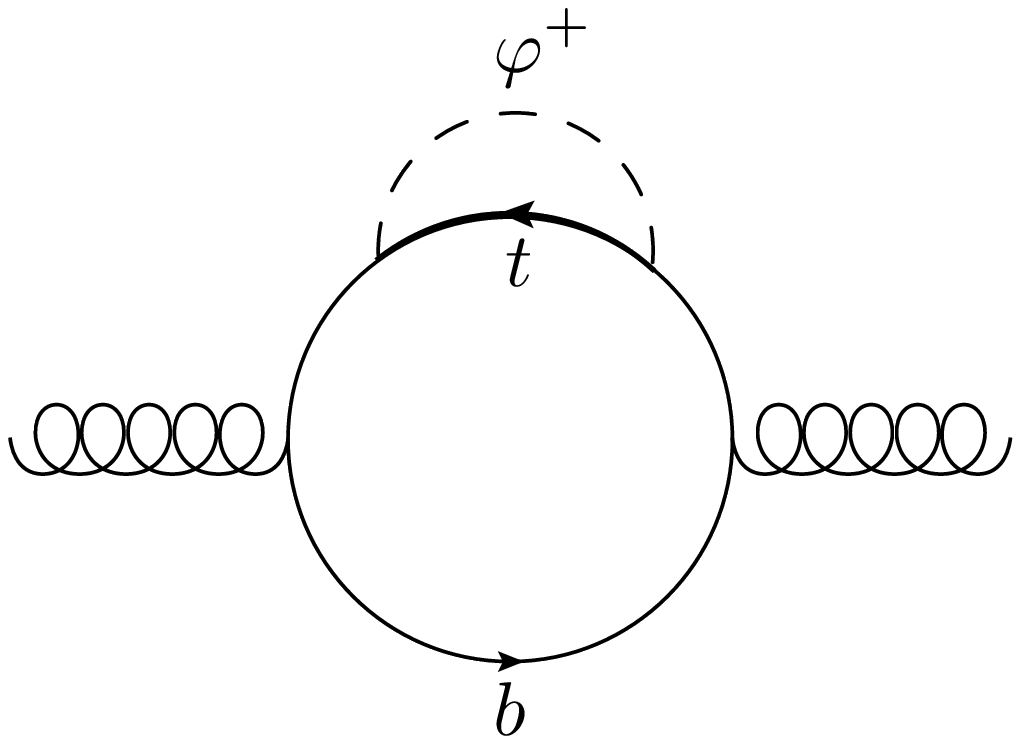} 
\end{center}
	&
	\begin{tabular}{c}
		${\cal F}_t \left[ -1 - \frac{1}{2 \ep} 
		-  \frac{p^2}{m_t^2} \left( \frac{ 4 \pi^2- 9}{96} +\frac{1}{16 \ep} \right.
		\right.$
		 \\
		$	\left. \left. + \frac{1}{8 \ep^2}\right)  - \frac{1}{18}\frac{p^4}{m_t^4} \right] $
		 \\
	\end{tabular}
&
	\begin{tabular}{c}
		${\cal F}_t \left[ -\frac{5}{16} - \frac{1}{8 \ep} 
		+  \frac{p^2}{m_t^2} \left(\frac{11}{288} + \frac{1}{48 \ep}\right) \right.$\\
		$ \left. + \frac{1}{144}  \frac{p^4}{m_t^4} \right] $	
	\end{tabular}
\\ \hline
\end{tabular*}
\end{center}
\caption{ 
Individual results for the Standard Model contractions. There is a prefactor 
\mbox{$
	{\cal F}_t = - \delta^{ab} (g^{\mu\nu}-p^\mu p^\nu/p^2) \frac{\alpha_s}{16 \pi^3} [N]^2 m_t^{2-4\ep} y_t^2
	$\; , with $y_t=m_t/v$.
}
}
\label{tab:individual_tensor_contractions_SM}
\end{table}
The diagrams where the bosons ($H,\phi^\pm,\phi^0$)  propagate on a leg 
(rows 2, 4, 6, and 7 of Table~\ref{tab:individual_tensor_contractions_SM}) have a symmetry factor of 2.  
The sum of the entries in Table~\ref{tab:individual_tensor_contractions_SM} is gauge invariant 
and gives the Standard Model result  
\bea
	\Pi^{\mu \nu, 2L}_{AB} \mid_{SM} 
	&=& 
	\frac{\alpha_s}{16 \pi^3} \delta_{AB} ( g^{\mu \nu} p^2 - p^\mu p^\nu)
		[N]^2 \frac{y_t^2}{3} m_t^{-4 \ep} \;,
		\label{eq:Pigg_SM}
\end{eqnarray}
with $[N]$ defined according to Eq.~\ref{eq:normNQ}. This result is finite and therefore,
using the low energy theorem of the previous Section,  
the two--loop ${\cal O}(y_t^3)$ 
contribution to the $ggH$ amplitude  is of order ${\cal O}(\ep)$. 

The terms needed for the renormalization of the one--loop amplitude are~\cite{Dawson:1988th,Djouadi:1997rj,Kniehl:1994ju}
\begin{eqnarray}
 	{\delta m_t\over m_t}\mid_{SM}&=&
 %{m_t^2\over 32 \pi^2 v^2}
	{y_t^2\over 32 \pi^2}[N]m_t^{-2\epsilon}
	\left({3\over \epsilon}+8\right) \;, \nn
	{\delta Z_{H}\over 2}\mid_{SM}&=&
 	-{N_C\over 16\pi^2} {y_t^2} [N]m_t^{-2\epsilon}\left[{1\over \epsilon}-{2\over 3} + {\cal O}(\ep^2) \right]
	 \; , \nn
	 {\delta M_W^2\over M_W^2}\mid_{SM}&=& 
	 - {N_C\over 8 \pi^2}{m_t^2\over v^2}
[N]m_t^{-2\epsilon}
	\left({1\over\epsilon}+{1\over 2}\right)
	\;,
\label{eq:renorm_sm}
\end{eqnarray}
where $N_C=3$.
Therefore, from Eq.~\ref{d3def},
\beq
\label{eq:delta3_sm}
	\delta_{3}\mid_{SM}= {7 \over 6} {N_C \over 16 \pi^2} {m_t^2\over v^2}  \;,
\eeq 
and the final two--loop ${\cal O}(y_t^3)$ contribution is, 
\begin{eqnarray}
	{\cal A}^{2L}_{gg \to H}\mid_{SM}  = 
	{\cal A}^{2L, \,ct}_{gg \to H}\mid_{SM} &=& {m_t^2 \over 16 \pi^2 v^2} 
		\left[ \frac{7}{6} N_C -3 \right] {\cal A}^{1L}_{gg \to H}\mid_{SM} 
		\nonumber \\
		&=& 0.0016\biggl({m_t\over 173~{\rm GeV}}\biggr)^2 {\cal A}^{1L}_{gg \to H}\mid_{SM} \; ,
\label{SMans}
\end{eqnarray}
with 
\bea
	{\cal A}^{1L}_{gg \to H}\mid_{SM} &=&
	- {\alpha_s\over 3 \pi v} 
	\delta_{AB} \left(g^{\mu\nu} p^2-p^{\mu} p^{\nu}\right) \;.
\label{eq:ggH1LSM}
\eea
Eq.~\ref{SMans} agrees with the results of Refs.~\cite{Djouadi:1994ge, Djouadi:1997rj}. 
%Finally, 
%\beq
%	\gamma_m \mid_{SM} = - {3 \over 16 \pi^2} y_t^2 
%	= {\rm finite} \left[-2 \ep \left( {\delta m_t \over m_t}\mid_{SM} \right) \right] \;,
%\eel
%as we discussed at the end of Sec.~\ref{sec:LET}.   ****MORE DETAILS HERE****

The results of Eq.~\ref{SMans} are to be compared with the total electroweak contribution to 
$gg\rightarrow H$~\cite{Djouadi:1994ge,Actis:2008ug,Aglietti:2004nj,Degrassi:2004mx,Actis:2008ts}.  
Assuming that the QCD and EW interactions factorize~\cite{Anastasiou:2008tj}, the electroweak
effects increase the total cross section by $\sim 5\%$ at the LHC~\cite{Dittmaier:2011ti}. 
The dominant role is played by light-fermion loops. The contribution from the top quark, also 
beyond the infinite--mass approximation, is just a few $\%$ of the light--quark 
contribution~\cite{Degrassi:2004mx}.
In order for the ${\cal O}(y_t^3)$ 
contributions to the cross
section to be ${\cal{O}}(5\%)$, we would have required $m_t\sim 700$~GeV, suggesting that in models
with heavy fermions the two--loop Yukawa  corrections might be the dominant electroweak
contribution~\cite{Passarino:2011kv}.  We will examine this possibility in the following Section.

\section{Top Partner Singlet Model}

\label{sec_basics}
\subsection{The Model}
We consider a model with an additional vector--like charge ${2\over 3}$ quark,  ${\cal{T}}^2$,  
which mixes with the Standard Model top quark~\cite{Dawson:2012mk, 
Dawson:2012di,Lavoura:1992qd,Aguilar-Saavedra:2013qpa, 
AguilarSaavedra:2002kr,Popovic:2000dx,Fajfer:2013wca}.  

For simplicity we make the following assumptions:
\begin{itemize}
\item The electroweak gauge group is the standard $SU(2)_L\times U(1)_Y$ group.
\item There is only a single Standard Model Higgs $SU(2)_L$ doublet, $\Phi$.

\item We neglect generalized CKM mixing and only allow mixing 
between the Standard Model -like top quark and 
the new charge ${2\over 3}$  singlet quark. 
\end{itemize}

The Standard Model--like fermions are,
\begin{equation}
\psi^1_L=\left(\begin{matrix}
{\cal{T}}_L^1\\
b_L \end{matrix}\right), \quad {\cal{T}}^1_R, b_R\, ,
\end{equation}
with the Lagrangian describing fermion masses,
\begin{equation}
-{\cal L}_M^{SM}=\lambda_1{\overline{\psi}}^1_L \Phi b_R+
\lambda_2{\overline{\psi}}^1_L {\tilde \Phi} {\cal{T}}^1_R+h.c. \; ,
\end{equation}
where ${\tilde \Phi}=i\sigma_2 \Phi^*$.  After electroweak symmetry breaking, 
the Higgs field is given by,
\begin{equation}
\Phi=\left(
\begin{matrix}
\phi^+\\
{1\over \sqrt{2}}(H+v-i\phi^0)
\end{matrix}\right) \; .
\end{equation}
Note that regardless of the Yukawa couplings, the Higgs boson and the neutral Goldston boson always
enter in the combination $H-i\phi^0$.

The mass eigenstates  are $t,T$ and $b$,
where $t$ and $b$ are the observed top and bottom quarks. 
The mass eigenstates in the top sector can be found by
the rotations: 
\begin{equation}
\chi_L^t\equiv
\left( \begin{matrix} 
t_L\\
T_L\end{matrix}\right)
\equiv  U_L^t
\left(\begin{matrix}
{\cal{T}}^1_L\\{\cal{T}}^2_L
\end{matrix}
\right)\, ,
\end{equation}
with   $\Psi_{L,R}\equiv{1\pm\gamma_5\over 2}\Psi$. Similar rotations are introduced 
for the right--handed fermions.  
The matrices $U_L^t$ and $U_R^t$  are unitary matrices 
and are parameterized as,
\begin{eqnarray}
U_L^t&=&
\left(\begin{matrix}
\cos\theta_L& -\sin\theta_L\\
\sin\theta_L& \cos\theta_L\end{matrix}
\right),\quad 
U_R^t=
\left(\begin{matrix}
\cos\theta_R & -\sin\theta_R\\
\sin\theta_R & \cos\theta_R\end{matrix}
\right) \, .
\end{eqnarray}
The most general CP conserving fermion mass terms
allowed by the $SU(2)_L\times U(1)_Y$ gauge symmetry 
 are,
\begin{eqnarray}
-{\cal L}_{M}&=&-{\cal L}_M^{SM}+
\lambda_3 
{\overline {\psi}}^1_L
{\tilde {\Phi}} 
{\cal {T}}^2_R+
\lambda_4
{\overline{{\cal {T}}}}^2_L 
{\cal {T}}^1_R+\lambda_5
{\overline{{\cal{T}}}}^2_L
{\cal {T}}^2_R+h.c.
\nonumber \\
&=&{\overline {\chi}}_L^t \biggl[U_L^t M^t U_R^{t \dagger}\biggr]\chi_R^t +
\lambda_1{v\over\sqrt{2}}{\overline{b}}_L b_R+h.c. \; ,
\end{eqnarray}
where
\begin{equation}
M^t=\left(
\begin{matrix}\lambda_2{v\over\sqrt{2}}&\lambda_3{v\over\sqrt{2}}\\
\lambda_4&\lambda_5\end{matrix}
\right) \; .
\end{equation}
We can always rotate ${\cal T}^2$ such that $\lambda_4=0$ 
and so there are 3 independent
parameters in the top sector, which we take to be the physical masses, $m_t$ and $M_T$, 
along with the left mixing angle, $\theta_L$. In the following we will abbreviate $s_L\equiv \sin\theta_L$, 
$c_L\equiv \cos\theta_L$. 
%The right-handed mixing angle is suppressed by a power of the large 
%mass, $\sin\theta_R={m_t\over M_T}\sin\theta_L$.

The couplings of the heavy charge ${2\over 3}$ quarks 
to the Higgs boson are~\cite{Dawson:2012di},
\begin{eqnarray}
-{\cal{L}}_H&=&{m_t\over v}c_L^2{\overline t}_Lt_RH
+{M_T\over v}s_L^2{\overline T}_LT_RH
+s_Lc_L {M_T\over v} {\overline t}_L T_RH
+s_Lc_L {m_t\over v} {\overline T}_Lt_RH
+h.c.\nonumber \\
&=&{m_t\over v} c_L^2{\overline t}tH
+{M_T\over v}s_L^2{\overline T}TH
+s_Lc_L{M_T+m_t\over 2 v}
\biggl( {\overline{t}} T+{\overline {T}}t\biggr)H
\nonumber \\ &&
+s_Lc_L\biggl({M_T-m_t\over 2 v}\biggr)
\biggl( {\overline{t}} \gamma_5 T-{\overline {T}}\gamma_5t\biggr)H\, .
\label{higgscoups}
\end{eqnarray}

The charged current interactions are,
\begin{eqnarray}
\label{eq:Wcouplings_S}
{\cal L}^{CC}
&=&-{g\over \sqrt{2}}
\biggl(
c_L 
{\overline{t}}_L
\gamma_\mu 
b_L+s_L
{\overline{T}}_L\gamma_\mu b_L\biggr)W^\mu
+h.c.\, .
\end{eqnarray}
Finally, the neutral current interactions are,
\begin{eqnarray}
{\cal L}^{NC}&=&{g\over \cos\theta_W}
\Sigma_{i=t,T}
\biggl\{\overline{f_i}\gamma^\mu
\biggl[(g_L^i+\delta g_L^i)\biggl({1-\gamma_5\over 2}\biggr)
+
(g_R^i+\delta g_R^i)\biggl({1+\gamma_5\over 2}\biggr)\biggr]
f_i
\biggr\}Z_\mu
\nonumber \\
&&+
{g\over \cos\theta_W}\Sigma_{i\ne j}
\biggl\{\overline{f_i}\gamma^\mu
\biggl[\delta g_L^{ij}\biggl({1-\gamma_5\over 2}\biggr)
+\delta g_R^{ij}\biggl({1+\gamma_5\over 2}\biggr)\biggr]f_j\biggr\}Z_\mu \, ,
\end{eqnarray}
where $g_L^i=T_3^i-Q_i s_W^2$, $g_R^i =-Q_i s^2_W$, $s_W$ is the sine of 
the Weinberg angle, $Q_i$ the electric charge of the quark  and
$T_3^i=\pm{1\over 2}$. The anomalous couplings are
\begin{eqnarray}
\delta g_L^t&=&\delta g_L^T=-{s_L^2\over 2} \;,\nonumber \\
\delta g_R^t&=&\delta g_L^T=\delta g_R^{tT}=0 \;, \nonumber \\
\delta g _L^{tT}& =&{s_Lc_L\over 2} \, .
\end{eqnarray}

\begin{figure}
\begin{center}
\includegraphics[bb=14 38 509 390,scale=0.6]{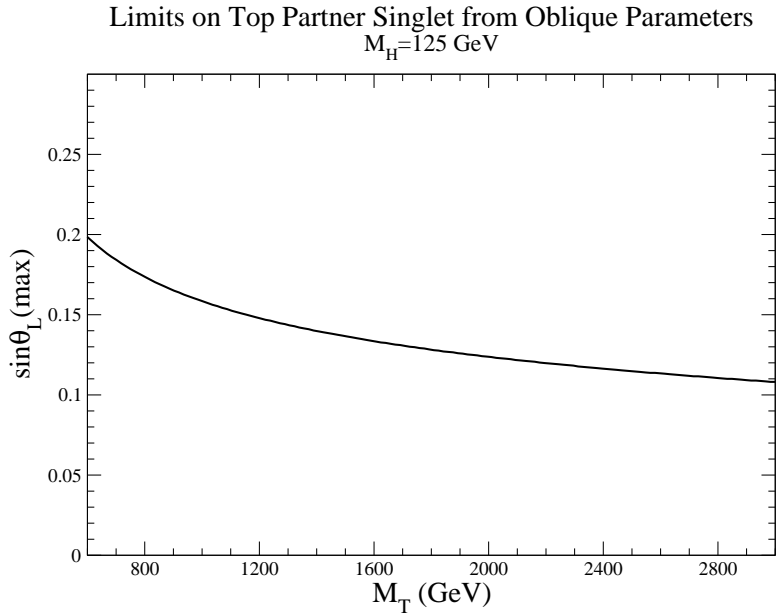} 
\vskip .25in 
\caption[]{Maximum allowed mixing angle, $\sin\theta_L$,  in the singlet top partner 
model from oblique parameters~\cite{Dawson:2012di}.}
\label{stu_sin}
\end{center}
\end{figure}

It is straightforward to use the above expressions to calculate the contributions of the
top partners to the oblique parameters, $\Delta S, \Delta T$ and 
$\Delta U$, and to parameters measured in 
$Z\rightarrow b {\overline b}$~\cite{Dawson:2012di,Aguilar-Saavedra:2013qpa,Fajfer:2013wca}. 
The most stringent restrictions are found from the oblique parameters
and are shown in Fig.~\ref{stu_sin}.
In the limit $M_T \sim m_t \gg M_W$, the top partner contributions to the $T$ 
parameter are~\cite{Dawson:2012di,Bai:2011aa},
\begin{equation}
\Delta {T}\sim {3\over 16\pi\sin^2\theta_W}
\biggl({M_T^2-m_t^2\over M_W^2}\biggr) s_L^2\, .
\label{dtlims}
\end{equation}
A scan over parameter space in the top singlet model~\cite{Dawson:2012di} 
using the exact results for $\Delta S$, $\Delta T$ and $\Delta U$ confirms
the accuracy of the approximate relationship of Eq.~\ref{dtlims} in the experimentally allowed
region.
It is clear that the heavy $T$ contributions decouple in the limit $s_L\rightarrow 0$.
Comparison with Eq.~\ref{higgscoups} shows that the mixed ${\overline{t}}\gamma_5 T H$ 
pseudoscalar couplings of the Higgs to top partners
are proportional to $\Delta {T}$ and hence must be highly suppressed.  We therefore neglect
these pseudoscalar couplings in the next Section.  We also note that the $T$ particle can be very heavy 
without being restricted by  the requirement of  perturbative unitarity in $F{\overline F}
\rightarrow F {\overline F}$ scattering~\cite{Chanowitz:1978mv, Dawson:2010jx},
\begin{equation}
s_L^2M_T < 550~{\rm GeV}\qquad {\hbox{(unitarity~bound)}} \, .
\end{equation}
For example $M_T=$2~TeV requires only $s_L<0.5 $ to preserve unitarity.

\begin{figure}
\begin{center}
\includegraphics[bb=14 38 509 390,scale=0.6]{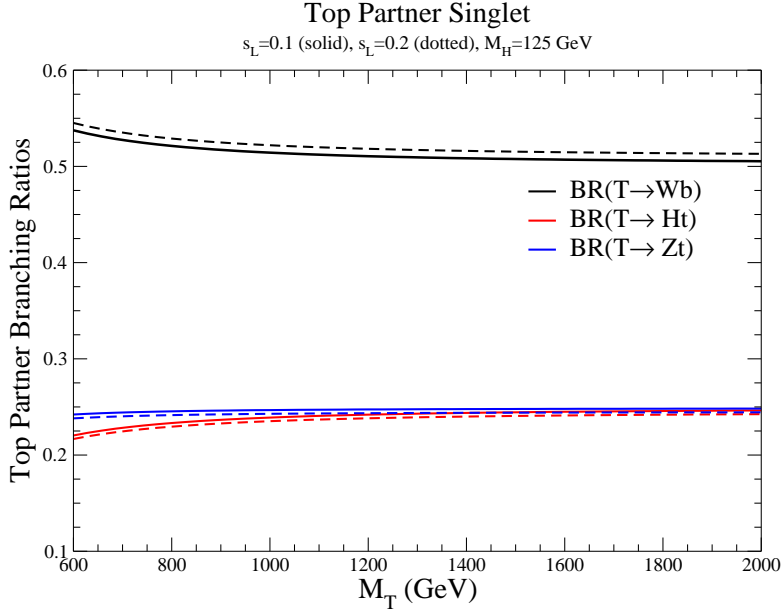} 
\vskip .25in 
\caption[]{Branching ratios of the top partner, $T$, in the singlet top partner model.}
\label{br_sin}
\end{center}
\end{figure}

Limits on the direct production of the top partner have been obtained by CMS~\cite{cmstop} as a function of the branching ratios, 
$T\rightarrow W^+b$, $T\rightarrow Zt$, 
and $T\rightarrow H t$.  These branching ratios are easily computed and are shown in Fig.~\ref{br_sin},
\begin{eqnarray}
\Gamma(T\rightarrow W^+ b)&=& {G_F\over 8 \pi \sqrt{2}}M_T\lambda^{1/2}(M_T,m_b,M_W)s_L^2(1+x_W^2-2x_W^4)
\nonumber \\
\Gamma(T\rightarrow Zt)&=& {G_F\over 16 \pi \sqrt{2}}M_T\lambda^{1/2}(M_T,m_t,M_Z)s_L^2c_L^2(1+x_Z^2-2x_t^2-2x_Z^4+x_t^4+x_Z^2x_t^2)
\nonumber \\
\Gamma(T\rightarrow Ht)&=& {G_F\over 16 \pi \sqrt{2}}M_T\lambda^{1/2}(M_T,m_t,M_H)s_L^2c_L^2(1+6x_t^2-x_H^2+x_t^4-x_t^2x_H^2)
\end{eqnarray}
where $\lambda(a,b,c)=a^4+b^4+c^4-2(a^2b^2+a^2c^2+b^2c^2)$, $x_i={M_i\over M_T}$, 
and we neglect
the $b$ mass.  The results are rather insensitive to $s_L$, as is obvious from Fig.~\ref{br_sin}.

In the following Sections, we compute the two--loop Yukawa enhanced contribution 
to $gg\rightarrow H$ in the
top partner singlet model using the low energy theorem.

\subsection{Contributions from off-diagonal terms}
\begin{figure}[t]
    \centering
   	\includegraphics[width=.8\textwidth]{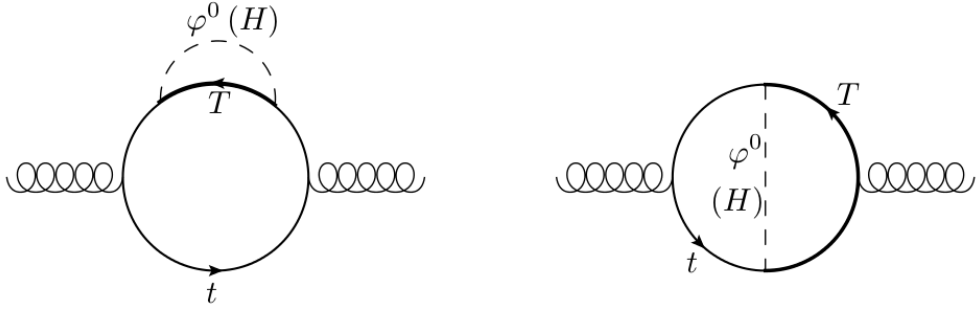} 
	\caption{Two-loop contributions to the gluon self-energy from ``mixed" diagrams.}
\label{fig:ggHmix}
\end{figure}

We first present results for the two--loop corrections to the gluon self-energy 
coming from diagrams involving two heavy quarks, $T$ and $t$, and a neutral 
boson, either $H$ or $\phi^0$. Examples of such diagrams are shown in 
Fig.~\ref{fig:ggHmix}.
We consider a general interaction Lagrangian
\bea
\label{eq:tTHphi_gen}
	- {\cal L}_{\Phi}^{N.C.} &=&
	(\overline{t}_L y_{tt} t_R + \overline{t}_L y_{tT} T_R 
	+ \overline{T}_L y_{TT} T_R  + \overline{T}_L y_{Tt} t_R) (H+i \phi^0) + h.c. \nn
	&=& 
	\phantom{+} H \left[  \overline{t} Y_{tt} t + \overline{t} Y_{tT} T 
		+\overline{T} Y_{TT} T  +\overline{T} Y_{Tt} t 
		+ \gamma_5 \left( \overline{t} A_{tT} T +\overline{T} A_{Tt} t \right) \right] \nn
	& & + i \phi^0
		\left[ \overline{t} A_{tT} T +\overline{T} A_{Tt} t + \gamma_5 
		\left( \overline{t} Y_{tT} T +\overline{T} Y_{Tt} t \right) \right] \; ,
\eea
where the couplings are assumed real. We defined
\beq
	Y_{qq'} ={ y_{qq'}+y_{q'q} \over 2}\, , \qquad \; A_{qq'} ={ y_{qq'} -y_{q'q} \over 2} \;.
\eeq
The diagonal interactions are pure scalar, as in the Standard Model. 
The corresponding contributions to the two--loop gluon self-energy can 
be obtained by rescaling the results of Table~\ref{tab:individual_tensor_contractions_SM}. 
We report here the contributions from the off--diagonal terms and the corresponding 
effects on the $ggH$ interaction in terms of the general Lagrangian, Eq.~\ref{eq:tTHphi_gen}. 
We will then adapt them to the top partner singlet 
model. 

The two--loop mixed diagrams with two different quarks and a Higgs boson exchange 
contribute,
\begin{eqnarray}
	\Pi^{\mu\nu,2L}_{AB}
%	\mid_{\begin{subarray}{l} mixed, \\ H \end{subarray}}&=&
	\mid_{mixed, H}&=&
	{\alpha_s \over 192 \pi^3} \delta_{AB} 
	\left(g^{\mu \nu} p^2 - p^\mu p^\nu \right) [N]^2
	m_t^{-4 \ep}
	\biggl[
		{1\over \ep} \left( \Delta_- + 4 \frac{a^2+1}{a} \Delta_+ \right)
\nn & & \;
		+ \frac{5}{2} \Delta_- + 4 \frac{a^2+1}{a} \Delta_{+}
		+ 4 \frac{\log a}{a^2-1} \left( \Delta_- -\frac{2 a^4-a^2-2}{a} \Delta_+ \right)\biggr] \; ,
%\nn	&=&
%	{\alpha_s \over 384 \pi^3} \delta_{AB} 
%	\left(g^{\mu \nu} p^2 - p^\mu p^\nu \right) [N]^2
%	\left({1 \over m_t M_T}\right)^{2 \ep}
%\nn &&
%	\times \left\{	
%		{2 \over \ep} \left[4 \left( {m_t \over M_T} + {M_T \over m_t} \right) (Y_{tT} Y_{Tt} + A_{tT} A_{Tt})
%		+ Y_{tT} Y_{Tt} - A_{tT} A_{Tt} \right] \right. \nn
%		& & \quad
%		+2 \log \left({m_t^2 \over M_T^2} \right) 
%		\left[ Y_{tT} Y_{Tt} {m_t+M_T \over m_t-M_T} - A_{tT} A_{Tt} {m_t-M_T \over m_t+M_T} \right] \nn
%		& & \quad \left. 
%		+8  {m_t^2+M_T^2 \over m_t M_T} (Y_{tT} Y_{Tt} + A_{tT} A_{Tt})
%		+5 (Y_{tT} Y_{Tt} - A_{tT} A_{Tt}) \right\} \; .
\end{eqnarray}
where we introduced the shorthand notation $a=M_T/m_t$, $\Delta_+ = Y_{tT} Y_{Tt} + A_{tT} A_{Tt}$, 
and
\mbox{$\Delta_- = Y_{tT} Y_{Tt} - A_{tT} A_{Tt}$}. 
This result correctly reproduces the limit for $M_T \to m_t$ of 
Table~\ref{tab:individual_tensor_contractions_SM} for zero pseudoscalar couplings and
$Y_{tT} = Y_{Tt} \rightarrow m_t Y_t/v$.
From the Lagrangian of Eq.~\ref{eq:tTHphi_gen}, the contribution from  the mixed diagrams with 
the exchange of a neutral Goldstone boson is 
\beq
	\Pi^{\mu\nu,2L}_{AB} 	\mid_{mixed, \phi^0} =
	- \Pi^{\mu\nu,2L}_{AB} 	\mid_{mixed, H}\left( Y_{qq'} \leftrightarrow A_{qq'} \right) \;,
\eeq
so that in the sum the terms which are symmetric under the exchange of $T$ and $t$ 
cancel and the total result from the mixed diagrams of Fig. \ref{fig:ggHmix}  is, 
\bea
\label{eq:PiggTt}
	\Pi^{\mu\nu,2L}_{AB} 	\mid_{mixed} & = &
	{\alpha_s \over 96 \pi^3} \delta_{AB} 
	\left(g^{\mu \nu} p^2 - p^\mu p^\nu \right) [N]^2
	m_t^{-4 \ep} \Delta_-
	\left({1 \over \ep} + \frac{5}{2} + 4 \frac{\log a}{a^2-1} \right)  \; .
%\nn &&
%		{\alpha_s \over 96 \pi^3} \delta_{AB} 
%	\left(g^{\mu \nu} p^2 - p^\mu p^\nu \right) [N]^2
%	\left({1 \over m_t M_T}\right)^{2 \ep}
%\nn &&
%	\times (Y_{tT} Y_{Tt} - A_{tT} A_{Tt})\left[
%		{1 \over \ep} +{m_t^2+M_T^2 \over m_t^2-M_T^2} \log \frac{m_t^2}{M_T^2} +\frac{5}{2}
%		\right] \;.
\eea
Again, this correctly reproduces the limit $M_T \to m_t$ from the sum of the first four 
entries of Table~\ref{tab:individual_tensor_contractions_SM} (accounting for a factor 
of 2 for the two heavy quarks and $(3-2 \ep)^{-1}$ for the projector of Eq.~\ref{eq:projdef}). 

Applying the low energy theorem, Eq.~\ref{eq:ggHLET}, the scalar $ggH$ vertex receives a 
finite correction 
\bea
\label{eq:Agg_Tt}
	{\cal A}^{0, 2L}_{gg \to H} \mid_{mixed} &=&
			{\alpha_s \over 24 \pi^3 v} \delta_{AB} 
	\left(g^{\mu \nu} p^2 - p^\mu p^\nu \right) \Delta_-
	\frac{1}{a^2-1}
	\left[ Y_T - a^2 Y_t -2 (Y_T-Y_t) \frac{a^2}{a^2-1} \log a \right] \;. \nn
%\nn &&			
%	{\alpha_s \over 24 \pi^3 v} \delta_{AB} 
%	\left(g^{\mu \nu} p^2 - p^\mu p^\nu \right) 
%	{ Y_{tT} Y_{Tt} - A_{tT} A_{Tt} \over m_t^2-M_T^2}
%	\nn && 
%	\times	\left[ Y_t M_T^2 - Y_T m_t^2 - (Y_t-Y_T) {m_t^2 M_T^2 \over m_t^2-M_T^2}
%	\log \frac{m_t^2}{M_T^2} \right]
%	\; .
\eea

\begin{figure}[t]
    \centering
   	\includegraphics[width=.7\textwidth]{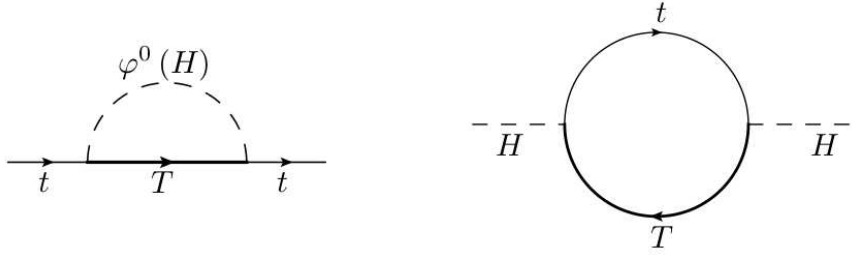} 
	\caption{Off--diagonal contributions to the quark (left) and Higgs (right) self 
	energy.}
\label{fig:renorm_s_offdiag}
\end{figure}
The off--diagonal couplings yield new contributions to the renormalization of the 
quark mass and of the Higgs wave function (Fig.~\ref{fig:renorm_s_offdiag}), 
\bea
\label{eq:trenorm_mix}
	{\delta m_t \over m_t}\mid_{mixed} &=&
	\frac{1}{16 \pi^2} [N] m_t^{-2 \ep} \Delta_- \left[
		{1 \over \ep} + 2 + a^2 - 2 a^4 \log a + (a^4-1) \log |a^2-1| \right] \; ,
\nn
	{\delta Z_H \over 2}\mid_{mixed}  &=&
		-\frac{N_C}{8 \pi^2} [N] m_t^{-2 \ep} 
	\biggl[
		\frac{\Delta_-}{\ep} + {\Delta_- \over 2} \frac{a^6-7a^4+7a^2-1-4a^4(a^2-3)\log a}{(a^2-1)^3} 
\nn && \;
	-\Delta_+ a \frac{a^4-1-4 a^2 \log a}{(a^2-1)^3} 
		\biggr] \;.
%\nn & &
%	-\frac{N_C}{8 \pi^2} [N] m_t^{-2 \ep} 
%	\biggl\{
%		\frac{Y_{tT} Y_{Tt} - A_{tT} A_{Tt}}{\ep} 
%		+ {1 \over 2} \frac{1}{(a^2-1)^2}
%\nn &&
%		\times \left[ (a+1)^2 (a^2-4a+1) Y_{tT} Y_{Tt} - (a-1)^2 (a^2+4a+1)  A_{tT} A_{Tt} \right]
%\nn &&
%		- {2 a^3 \log a \over (a^2-1)^3} \left[ (a^3- 3 a - 2) Y_{tT} Y_{Tt} - (a^3 - 3 a +2) A_{tT} A_{Tt}
%		\right]		
%		 \biggl\} \;.
%
\eea
These results correctly reproduce the Standard Model limit, Eq.~\ref{eq:renorm_sm}, from the
top quark contribution when $\Delta_+=\Delta_-\rightarrow ({m_t^2\over v^2})$ and $a\rightarrow 1$.
% -- \ref{eq:delta3_sm}) 
(Note that for the quark mass renormalization one also 
needs to add the contribution from the $b$ quark loop, as in the Standard Model). 
The $W$ mass receives contributions from $t-b$ and $T-b$ loops. They have the same 
form as in Eq.~\ref{eq:renorm_sm} up to rescaling factors $V_{tb} V^{*}_{bt}$, $V_{Tb} V^{*}_{bT}$ 
from modifications of the heavy--quark couplings to the $W$ bosons with respect to the 
Standard Model. These modifications are related to the deviations in the $F {\overline F}'H$
vertex. In the singlet top partner  model we  consider, 
both come from the mixing among quarks of the same quantum 
numbers. 
Poles will cancel once we consider  an explicit model where these relations are clear. 
Since the two--loop amplitude, Eq.~\ref{eq:Agg_Tt}, is finite and the quark mass 
renormalization only contributes a finite term (Eq.~\ref{eq:2L_renorm}), we 
expect $\delta_3$ to be finite.

\subsection{Results for Top Partner Model}
We now turn to the top partner singlet model described at the beginning of this Section. 
The two--loop gluon self--energy containing only heavy quarks of one kind, either $t$
or $T$, is finite, as shown in Eq.~\ref{eq:Pigg_SM}. Therefore,  there is no contribution to the 
unrenormalized \mbox{two--loop amplitude},
${\cal A}_{gg \to H}^{0, 2L}$, coming from the diagonal fermion interactions
of Eq.~\ref{higgscoups}. 
Applying the couplings of Eq.~\ref{higgscoups} to the general result of Eq.~\ref{eq:trenorm_mix},
the diagrams containing off--diagonal mixings 
between different heavy quarks and the bosons inside the loop yield a contribution,\footnote{We
use the subscript `s' to denote quantities in the top singlet model.}
\bea
	\label{eq:Agg_singl}
	{\cal A}^{0, 2L}_{gg \to H} \mid_{s} &=& {\cal A}^{0, 2L}_{gg \to H} \mid_{mixed} \nn
	&=&			
	-{m_t^2 \over 16 \pi^2 v^2} %\delta_{AB} 
%	\left(g^{\mu \nu} p^2 - p^\mu p^\nu \right) 
	s_L^2 c_L^2 
	\frac{a^2+1}{a^2-1} \left[s_L^2-a^2 c_L^2 + (c_L^2-s_L^2) \frac{2 a^2}{a^2-1} \log a \right] 
	\times
		{\cal A}^{1L}_{gg \to H} \mid_s\;.
\eea
We normalized the result to the one--loop $ggH$ amplitude in the top  partner singlet model, 
\bea
	{\cal A}^{1L}_{gg \to H} \mid_s &=&
	-\frac{\alpha_s}{3 \pi v}
	 \delta_{AB} (g^{\mu \nu}p^2-p^{\mu} p^{\nu}) \;.
\eea
Note that in the infinite mass approximation, this is the same as the Standard Model 
amplitude, Eq.~\ref{eq:ggH1LSM}. Only finite mass corrections yield deviations
from the Standard Model result~\cite{Dawson:2012di}. 

The low energy theorem as formulated 
in Eq.~\ref{eq:ggHLET} does not reproduce the diagrams where the external Higgs boson couples 
to two different quarks.  From Eq.~\ref{higgscoups}, we see that these pseudo-scalar 
couplings are proportional to $s_L(M_T-m_t)\sim \Delta T$ and are restricted by the 
measurements shown in Fig.~\ref{stu_sin}  to be small.  Neglecting them is thus a reasonable approximation.

From Eq.~\ref{eq:Wcouplings_S}, the couplings of the heavy quarks to the 
$W$ boson in the singlet model
are rescaled by $V_{tb}=V_{bt}^{*} = c_L$, $V_{Tb}=V_{bT}^{*} = s_L$, and the 
$W-$mass renormalization is,
\bea
	{\delta M_W^2\over M_W^2}\mid_{s}&=&-{N_C\over 8 \pi^2 v^2}[N]
	\biggl({1\over \epsilon}+{1\over 2}\biggl) m_t^{2-2\epsilon}
	\biggl[c_L^2+s_L^2 a^2(1-4 \log a)\biggr] \, .
\label{wtop}
\eea

The wave function renormalization is found by rescaling the Standard Model results of 
Eq.~\ref{eq:renorm_sm} by $c_L^4$ for the $t$
contribution, $s_L^4$ for the $T$ contribution, and adding the mixed contribution of 
Eq.~\ref{eq:trenorm_mix} using
the couplings of Eq.~\ref{higgscoups},
\begin{eqnarray}
  {\delta Z_H\over 2}\mid_{s}&=&-{N_C\over 16\pi^2 v^2}[N] m_t^{2-2\ep}
  \biggl[
		\frac{a^2 s_L^2+c_L^2}{\ep}-\frac{2}{3} (a^2  s_L^4 + c_L^4)
		- 2a^2 \log a \, s_L^4
\nn & & \phantom{\biggl\{}
			+ s_L^2 c_L^2 \frac{
				a^8-10 a^6+10 a^2-1 -4 a^4 (a^4-2 a^2-7) \log a }{2 (a^2-1)^3}
  \biggr]\, .
  \label{zhtop}
\eea	
From the general result of Eq.~\ref{d3def}, we obtain
\bea
	\delta_3\mid_{s} &=&
	{N_C\over 96\pi^2 v^2} m_t^{2}
  \biggl[
  	4 (c_L^4 + s_L^4 a^2) +s_L^2 a^2 (3-12 c_L^2 \log a) + 
 \nn & & \;
  	3 c_L^2 \left(1- s_L^2 \frac{a^6-9a^4-9a^2+1}{(a^2-1)^2} \right)
  	+12 s_L^2 c_L^2 a^4 \log a \frac{a^4-2 a^2-7}{(a^2-1)^3} \biggr] \; .
\label{eq:delta2_s}
\eea
As we anticipated, the poles in the $W$ mass and the Higgs wave function 
renormalization cancel and $\delta_3$ is finite. In the limits $c_L \to 1$ or
$s_L \to 1$ only one heavy quark ($t$ or $T$, respectively) couples to the Higgs, 
and Eq.~\ref{eq:delta2_s} correctly reproduces the Standard Model infinite--mass 
result of Eq.~\ref{eq:delta3_sm}. 
The final ingredient that we need for the two--loop renormalization are the poles 
of the heavy quark mass renormalization constants. Combining Eqs.~\ref{eq:renorm_sm} 
and~\ref{eq:trenorm_mix},
\bea
	{\delta m_t \over m_t}\mid_{s,\ep} &=& \frac{1}{\ep}
	\frac{1}{32 \pi^2 v^2} m_t^{2} c_L^2 \left[
		3 c_L^2+s_L^2 (a^2+1) \right] \; , \nn
		{\delta M_T \over M_T}\mid_{s,\ep} &=& \frac{1}{\ep}
	\frac{1}{32 \pi^2 v^2} m_t^{2} s_L^2 \left[
		3 s_L^2 a^2+c_L^2 (a^2+1) \right] \; .
\eea	

From Eq.~\ref{eq:2L_renorm} the two--loop counterterm is,
\bea
		{\cal A}^{2L, \,ct}_{gg \to H}\mid_s &=&
		\frac{m_t^2}{64 \pi^2 v^2} \frac{1}{(a^2-1)^2} 
		\biggl[
		a^6+5 a^4+5 a^2 +1 
\nn &&
	\phantom{\times} - (a^2-1)^3 \frac{11  \cos (2 \theta_L) - 3 \cos (6 \theta_L)}{8}
		-6 a^2(a^2 +1) \cos (4 \theta_L) 
\nn && \phantom{\times}
		+6 a^2 \log a \frac{a^4-10a^2+1}{a^2-1} \sin^2(2 \theta_L) \biggr]  \times
		{\cal A}^{1L}_{gg \to H} \mid_s\;,
\eea
where for simplicity we set $N_C=3$.
The renormalized two--loop amplitude then reads
\bea
	{\cal A}^{2L}_{gg \to H}\mid_s
	&=&
	\frac{1}{256 \pi^2 v^2} {m_t^2 \over (a^2-1)^3} 
	\biggl\{
		5 a^8 + 14 a^6 -14 a^2-5 
\nn && 
			+8  \sin^2(2 \theta_L) a^2\log a \biggl[
			3(a^4-10 a^2+1) -(a^4-1) \cos(2 \theta_L) \biggr]
\nn &&
		-\cos(2 \theta_L) (a^2-1)^2 (5 a^4-12 a^2+5) 
		- \cos(4 \theta_L) (a^8+22 a^6-22 a^2-1)		
\nn && 
		+ \cos(6 \theta_L)	(a^2-1)^2 (a^4-4 a^2+1)
			\biggr\} \times
		{\cal A}^{1L}_{gg \to H} \mid_s\;.
\label{singans}			
\eea
This reproduces the Standard Model result of Eq.~\ref{SMans} 
for $\theta_L=0$ and in the $a\rightarrow 1$ limit for $\theta_L= \pi/2$, i.e., when only one heavy 
quark of mass $m_t$ runs in the loops with Standard Model--like couplings. 
For small mixing, as required
by the precision electroweak results,  Eq.~\ref{singans} reduces to,
\begin{eqnarray}
{\cal A}^{2L}_{gg \to H}\mid_s 
	&\! \rightarrow \! & 
\frac{m_t^2}{32 \pi^2 v^2} \biggl[ 1 + \frac{2 \theta_L^2}{(a^2-1)^2} 
   \biggl( 15 a^4+8 a^2+1 
  +  4 a^2 \log a  \frac{a^4-15 a^2+2}{a^2-1}  \biggr)
  \biggr] {\cal A}^{1L}_{gg \to H} \!\! \mid_s  \;. \nn
 \label{eq:AsSmallAngle}
\end{eqnarray}
In the limit of small $\delta \equiv M_T - m_t$, but for arbitrary mixing,
\bea
	{\cal A}^{2L}_{gg \to H}\!\!\mid_{s} 
	&\!\! \rightarrow \! & \!
	\frac{ m_t^2}{32 \pi^2 v^2}\biggl[
  3-2 \cos(4 \theta_L)  - \frac{\delta}{6 m_t} \sin^2 \theta_L
  \biggl(7\cos(4 \theta_L)-34 \cos(2 \theta_L)-53\biggr) \biggr] {\cal A}^{1L}_{gg \to H}\!\mid_s \;.\nn
\end{eqnarray}
%\nonumber \\
%&&	
%	\frac{\alpha_s}{\pi^3 v^4} {m_t^3}
%	\delta_{AB}(g^{\mu\nu}-p^\mu p^\nu)
%	 \left[
%		\cos(4 \theta_L)-{3 \over 2} \right] \;.
%\eea 
Finally, for almost degenerate quarks with small mixing both these expansions 
reduce to
\bea
	{\cal A}^{2L}_{gg \to H}\!\!\mid_{s} 
	&\! \rightarrow \! & 
	\frac{ m_t^2}{96 \pi^2 v^2}\left(
  3 + 48  \theta_L^2 +40  \frac{\delta}{m_t}  \theta_L^2\right) {\cal A}^{1L}_{gg \to H}\!\mid_s \;.
\end{eqnarray}
The first correction in the small $\frac{\delta}{m_t}$ parameter is further suppressed by the 
small mixing, and is therefore subleading with respect to the $\theta_L^2$ correction. 

In the $\delta\rightarrow 0$ limit there are no $t\gamma_5 {\overline t}H $ contributions, and the 
low energy theorem reproduces all contributions.

\subsection{Phenomenology}
\label{sec:phen}
\begin{figure}
\begin{center}
\includegraphics[bb=14 38 509 390,scale=0.6]{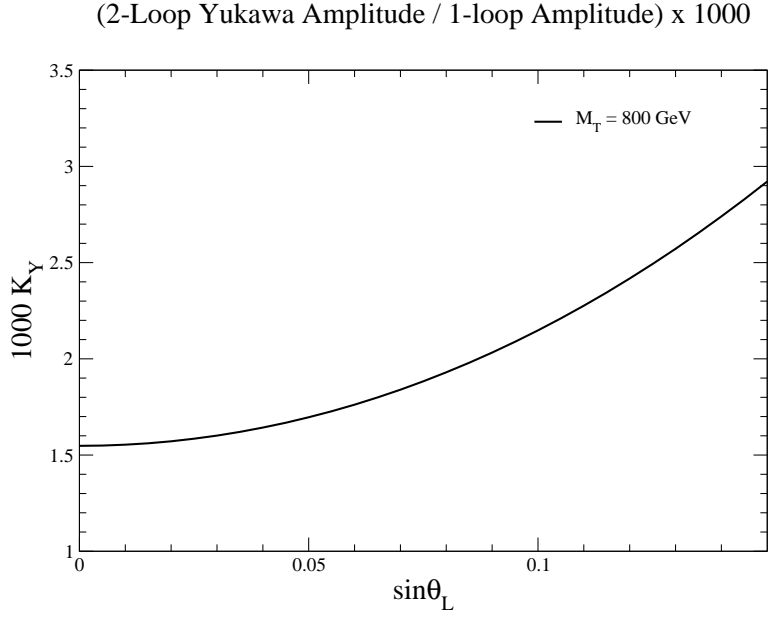} 
\vskip .25in 
\caption[]{1000 times the contribution from the two--loop Yukawa amplitude of
Eq.~\ref{singans} divided by the one--loop Yukawa amplitude in the Top Partner Singlet
Model with $M_T=800$~GeV.}
\label{sing_tl}
\end{center}
\end{figure}

\begin{figure}
\begin{center}
\includegraphics[bb=14 38 509 390,scale=0.6]{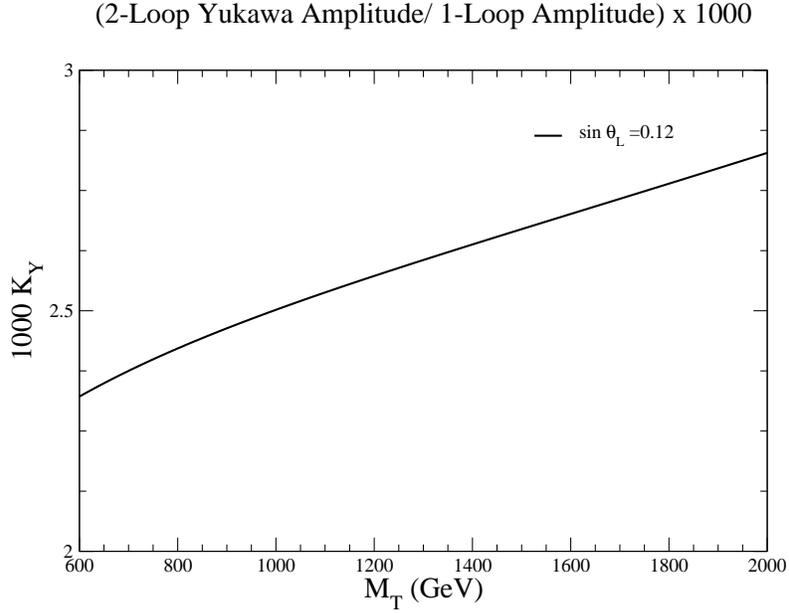} 
\vskip .25in 
\caption[]{1000 times the contribution from the two--loop Yukawa amplitude of
Eq.~\ref{singans} divided by the one--loop Yukawa amplitude in the Top Partner Singlet
Model with $\sin\theta_L=.12$.}
\label{sing_mt}
\end{center}
\end{figure}

In this Section, we consider the phenomenological implications of the Yukawa corrections
to the top partner model given in the previous Section. At one--loop, the amplitude
for $gg\rightarrow H$ is identical to the Standard Model rate up to corrections
of ${\cal O}({M_H^2\over M_T^2})$~\cite{Dawson:2012di}, and large deviations are therefore 
first possible at the two--loop level. 
\begin{figure}
\begin{center}
\includegraphics[bb=14 38 509 390,scale=0.6]{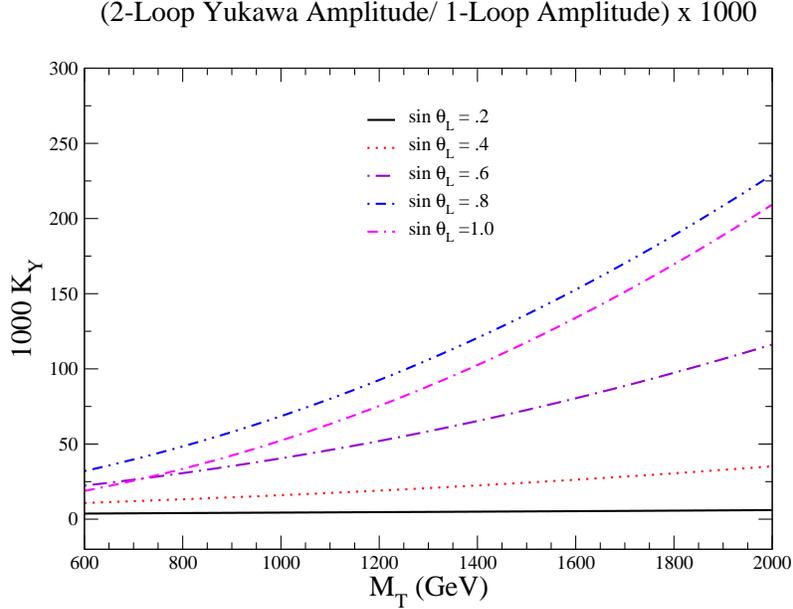} 
\vskip .25in 
\caption[]{1000 times the contribution from the two--loop Yukawa amplitude of
Eq.~\ref{singans} divided by the one--loop Yukawa amplitude in the Top Partner Singlet
Model as a function of the mixing with the Standard Model top quark.}
\label{big}
\end{center}
\end{figure}
In Fig.~\ref{big}, we show the effects
of the two--loop contributions relative to the one--loop contribution (including only the Yukawa
terms calculated here), without keeping into account the bounds from electroweak 
precision data. We quantify these effects through the $K-$factor
\beq
	K_Y  ={ {\cal A}^{2L}_{gg\to H}\mid_s \over  {\cal A}^{1L}_{gg\to H}\mid_s}\; .
\eeq
Only for ridiculously large values of the mixing parameter, $s_L\sim 1$,
do the effects of the Yukawa corrections reach the level of a few $\%$. 
Effects are even smaller if we restrict ourselves to the allowed region of Fig.~\ref{stu_sin} 
(Figs.~\ref{sing_tl} and~\ref{sing_mt}).
In Fig.~\ref{sing_tl}, we show how the Yukawa corrections increase with the mixing angle 
for a fixed $M_T=800$~GeV, in the range allowed by precision electroweak measurements. 
The behavior is consistent with the small--angle expansion of Eq.~\ref{eq:AsSmallAngle}. 
Finally, in Fig.~\ref{sing_mt}, we show the dependence on the heavy mass $M_T$. For 
large values of $M_T$ to be allowed, we need to restrict ourselves to a small mixing 
angle. It is clear that these two--loop Yukawa corrections
are always at the sub-percent level.  In order to obtain  large Yukawa corrections, we would need to construct
a more complicated model where large mixing with the Standard Model fermions 
was not forbidden by electroweak precision 
measurements.

\section{Conclusions}
\label{sec:conc}
We have considered the two--loop ${\cal O}\left(\left(\frac{Y_F M_F}{v}\right)^3\right)$ 
contributions to $gg\rightarrow H$ using the
low energy theorem and our analytic results will be of use to future model builders.
These corrections are well known and small for the Standard Model. In the singlet top partner model
there are contributions of ${\cal O}\left(\left(\frac{Y_F M_F}{v}\right)^3\right)$ 
to Higgs production via gluon fusion which
are potentially important.  These corrections are suppressed, however, by a mixing angle, $s_L$, 
which is restricted by precision electroweak measurements to be small, and we find that the 
Yukawa corrections
in this model are at the sub-precent level.  This reinforces our conclusions from a previous work, 
that the singlet top
partner model represents an example where the gluon fusion Higgs production rate
 will be almost identical to that of the Standard Model 
and hence precision measurements of the rate will be insensitive to the new physics.  Exploring this class of
models will require the direct observation of the top partners.

\section*{Acknowledgements}
BNL is supported by the U.S. Department of Energy under grant
No.~DE-AC02-98CH10886. 
Fermilab is operated by the Fermi Research Alliance under contract no. 
DE-AC02-07CH11359 with the U.S. Department of Energy.
EF thanks the Galileo Galilei Institute for the hospitality during the 
completion of this work.

%\newpage
\bibliographystyle{unsrt}
\bibliography{paper}

\end{document}